# Correlations of structural, magnetic, and dielectric properties of undoped and doped CaCu$_3$Ti$_4$O$_{12}$

S. Krohns,[1,a)] J. Lu,[1,2] P. Lunkenheimer,[1] V. Brizé,[2] C. Autret-Lambert,[3] M. Gervais,[3] F. Gervais,[3] F. Bourée,[4] F. Porcher,[4] and A. Loidl[1]

[1] *Experimental Physics V, Center for Electronic Correlations and Magnetism, University of Augsburg, 86135 Augsburg, Germany*
[2] *School of Materials Science and Engineering, University of Science and Technology Beijing, Beijing 100083, China*
[3] *Laboratoire LEMA, UMR 6157 CNRS-CEA, Université F. Rabelais, Parc de Grandmont, 37200 Tours, France*
[4] *Laboratoire Léon Brillouin CEA Saclay 91191, Gif Sur Yvette Cedex, France*

The present work reports synthesis, as well as a detailed and careful characterization of structural, magnetic, and dielectric properties of differently tempered undoped and doped CaCu$_3$Ti$_4$O$_{12}$ (CCTO) ceramics. For this purpose, neutron and x-ray powder diffraction, SQUID measurements, and dielectric spectroscopy have been performed. Mn-, Fe-, and Ni-doped CCTO ceramics were investigated in great detail to document the influence of low-level doping with 3d metals on the antiferromagnetic structure and dielectric properties. In the light of possible magnetoelectric coupling in these doped ceramics, the dielectric measurements were also carried out in external magnetic fields up to 7 T, showing a minor but significant dependence of the dielectric constant on the applied magnetic field. Undoped CCTO is well-known for its colossal dielectric constant in a broad frequency and temperature range. With the present extended characterization of doped as well as undoped CCTO, we want to address the question why doping with only 1% Mn or 0.5% Fe decreases the room-temperature dielectric constant of CCTO by a factor of ~100 with a concomitant reduction of the conductivity, whereas 0.5% Ni doping changes the dielectric properties only slightly. In addition, diffraction experiments and magnetic investigations were undertaken to check for possible correlations of the magnitude of the colossal dielectric constants with structural details or with magnetic properties like the magnetic ordering, the Curie-Weiss temperatures, or the paramagnetic moment. It is revealed, that while the magnetic ordering temperature and the effective moment of all investigated CCTO ceramics are rather similar, there is a dramatic influence of doping and tempering time on the Curie-Weiss constant.

## I. INTRODUCTION

A new class of functional materials based on perovskite-related compounds like the prominent CaCu$_3$Ti$_4$O$_{12}$,[1,2,3,4,5,6,7,8,9,10,11,12] has attracted considerable attention, stimulated by prospective applications involving the enhancement of the performance of capacitive electronic elements. This material class shows a number of interesting effects, like very high dielectric constants, which are enhanced in a broad temperature and frequency range,[2,3,13] ferrimagnetism with considerable ferromagnetic moments and high ordering temperatures,[14,15,16] or even heavy fermion-like behavior.[17,18] Especially materials with extremely high ("colossal") dielectric constants (CDCs) are urgently needed and there are many efforts to prepare high-quality thin films of CCTO as a first step to application of this material in modern electronics.[19] However, even the properties of bulk samples are not well understood so far and there are ongoing discussions where the CDCs in CCTO originate from. However, at least it seems now more or less commonly accepted that the CDC feature is due to extrinsic effects, like surface barrier layer capacitances[11,20,21,22] or internal barrier layer capacitances.[6,10,23] Besides the CDC, in the view of possible magnetoelectric coupling it also is important to carefully study the magnetic structure as well as the dielectric properties in external magnetic fields.[24]

It is well known that the dielectric properties of CCTO strongly depend on processing conditions during sample preparation, on oxygen non-stoichiometry, as well as on small amounts of doping. Concerning impurity effects, results reported in literature have demonstrated that doping with 3d cations leads to a systematic and significant decrease of the effective dielectric constant,[25,26] which is accompanied by a concomitant strong decrease in conductivity.[27,28] Especially, CCTO samples doped with marginal amounts of Fe or Mn with the nominal chemical composition CaCu$_3$Ti$_{4-x}$B$_x$O$_{12}$ revealed strong doping effects, both in the electron-paramagnetic response, namely the onset of ferromagnetic-type spin interaction below $T_N$, as well as in the dielectric properties, shifting the CDC relaxation time to higher temperatures.[29]

Most interestingly, the dc conductivity of CCTO becomes dramatically reduced when doping it with Mn.[27] The detected decrease of conductivity by several orders of magnitude directly influences the Maxwell-Wagner like relaxation and shifts the relaxation responsible for the

a)Electronic mail: stephan.krohns@physik.uni-augsburg.de

CDC to higher temperatures and lower frequencies. This fact allows the investigation of the intrinsic complex dielectric properties in a much broader temperature range.[30] In undoped CCTO it was demonstrated that the investigation of intrinsic dielectric properties in a broad temperature range is only possible in the GHz frequency regime and beyond.[31] In addition, manganese doped CCTO shows very interesting low temperature properties, which were explained by incipient ferroelectricity.[30]

For the present work, we have investigated a number of Mn-, Fe-, and Ni-doped CCTO samples. The samples were the same as those studied earlier by electron spin resonance.[29] In addition, undoped CCTO, subjected to different tempering times was investigated (same samples as studied in Ref. 32). The magnetic properties of all samples are reported. For undoped, Ni-, and Fe-doped CCTO, we provide a detailed structural analysis using neutron-powder diffraction. The dielectric properties of all samples were determined in detail, especially concentrating on Fe- and Mn-doped CCTO and also performing dielectric measurements in external magnetic fields. By these studies we specifically search for correlations between structural, magnetic, and dielectric properties, including possible magnetoelectric effects.

## II. EXPERIMENTAL DETAILS

CCTO ceramics doped with Ni, Fe, and Mn were prepared by an organic gel-assisted citrate synthesis[33] and sintering process. First, triammonia citrate as a chelating agent was added to an aqueous solution of metal (M) nitrates with M = calcium, copper, and, for the doped samples, iron, nickel, or manganese in appropriate amounts. Titanium citrate formed from titanium alcoxide was then added and a clear solution stable up to the gel pyrolysis was obtained. The solution was gelled by in situ formation of an auxiliary three-dimensional polymeric network. The monomers acrylamide and $N,N'$-methylenediacrylamide were dissolved and co-polymerized by heating at about 100°C with azobisisobutyronitrile (AIBN) as a radical polymerization initiator.[34] The aqueous gel was then calcined for 20 h at 500°C for CCTO and at 650°C for doped CCTO. The resulting powders were treated in two different ways according to the subsequent analyses. For structural characterizations they were heated at 1000°C for 20 h. For dielectric measurements the ceramics were ground, cold pressed into disks of 10 mm in diameter and 1 mm in thickness with polyvinyl alcohol binder and then sintered in air at 1000°C for 20 h.[23] After that treatment the samples were cut into smaller pieces (A ≈ 3×4 mm$^2$, d ≈ 1 mm). The differently tempered polycrystalline CCTO samples were prepared, as reported in Ref. 32.

For the dielectric measurements, silver paint contacts were applied to opposite sides of the plate-like samples. Dielectric spectroscopy was performed in a frequency range from 1 Hz up to 1 MHz and in a temperature range from 2 K up to 425 K using a Novocontrol Alpha-A Frequency-Response Analyzer and a Hagerling AH2700a autobalance bridge.[35] The applied ac voltage was 1 V. Dielectric spectroscopy on selected samples was also performed in external magnetic fields up to 7 T.

Magnetic susceptibilities were measured using a commercial SQUID. Structural characterization was performed with x-ray (BRUKER D8, Cu-K$_\alpha$ radiation) and neutron-powder diffraction (high resolution 3T2 diffractometer in LLB-Saclay, France, λ = 1.2261 Å). Elemental analysis was carried out using energy dispersive spectroscopy (EDS) coupled with a transmission electron microscope (JEOL 2100) applied to about 50 points on each sample.

## III. STRUCTURAL CHARACTERIZATION

Annealed powders of all samples as well as sintered disks were found to display single-phase CCTO x-ray diffraction patterns pertinent to the cubic CCTO structure. The composition as determined from EDS was revealed to be uniform. No secondary phases were detected, neither from diffractometry nor from EDS.

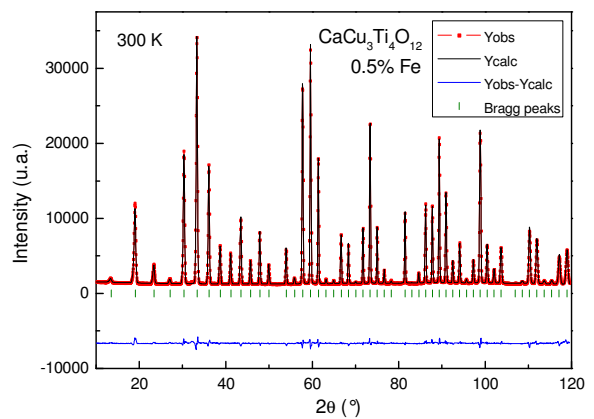

FIG. 1. (Color online) Experimental (circles), calculated (line), and difference (lower line) neutron powder diffraction patterns of CCTO doped with 0.5% Fe at room temperature (λ = 1.226 Å). The positions of Bragg peaks of the Im-3 phase are marked at the zero intensity line.

Undoped, Ni-, and Fe-doped CCTO powder samples were studied at room temperature by neutron-powder diffraction. The Rietveld method using the Fullprof program was employed to refine the structural and profile parameters.[36] A pseudo-Voigt function was chosen for the line shape of the diffraction peaks. Lattice parameters, positional coordinates, isotropic thermal, and occupancy factors were refined. The structural refinements from the room temperature high-resolution neutron-powder-diffraction data were performed in the space group Im-3, characterized by cell parameters that are doubled when compared to the unit cell of the ideal cubic perovskite structure, with Ca in 2a (0 0 0), Cu in 6b (0 ½ ½), Ti in 8c (0 ¼ ¼), and O in 24g (x y 0) sites. For CCTO the cell parameter a = 7.39746(3) Å. For Fe- or Ni-doped samples, the lattice constants do not change significantly. The best fits were obtained assuming 3d substitutions on the Ti site. All samples show oxygen deficiency exceeding the estimated errors, which increases from CCTO to the Fe-doped samples (see Table I). The resulting essential crystallographic information, namely lattice constants, atomic distances, bond-valence sums (BVS), and the reliability factors of the fits are listed in Table I. Refined diagrams for the Fe- and Ni-doped samples are shown in Figs. 1 and 2, respectively. In a broad range of scattering angles the Rietveld refinement works very well and no impurity phases above background level could be detected.

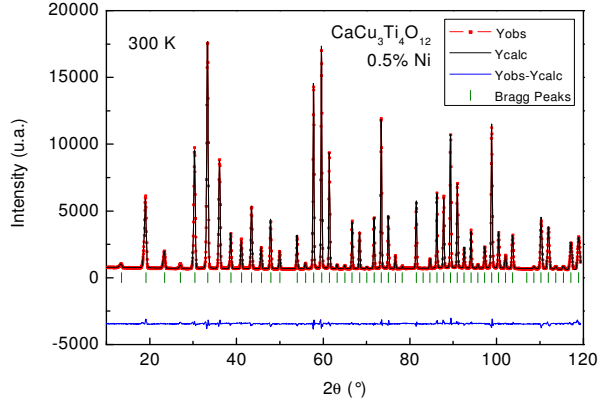

FIG. 2. (Color online) Experimental (circles), calculated (line), and difference (lower line) neutron powder diffraction patterns of CCTO doped with 0.5% Ni at room temperature ($\lambda = 1.226$ Å). The positions of the Bragg peaks of the Im-3 phase are marked at the zero intensity line.

|  | undoped | 0.5% Ni doped | 1% Fe doped |
|---|---|---|---|
| $R_{Bragg}$ | 1.73 | 1.99 | 2.38 |
| $\chi^2$ | 6.91 | 5.48 | 7.81 |
| Ca-O (Å) | 2.6056 (4) | 2.6061 (6) | 2.6076 (5) |
| Cu-O (Å) | 1.9713 (4) | 1.9705 (6) | 1.9702 (5) |
| Ti-O (Å) | 1.9611 (4) | 1.9612 (2) | 1.9676 (4) |
| BVS Ca | 2.108 (1) | 2.071 (1) | 2.096 (1) |
| BVS Cu | 1.792 (1) | 1.766 (1) | 1.796 (1) |
| BVS Ti | 3.990 (1) | 3.924 (2) | 3.983 (2) |
| BVS O | 1.969 (1) | 1.918 (1) | 2.052 (1) |

TABLE I. Reliability factors, inter atomic distances, and bond valence sums (BVS) for three samples: undoped CCTO, as well as 0.5% Fe and 0.5% Ni doped ceramics.

For one undoped (actual composition: $Ca_{0.98}Cu_{2.95}Ti_4O_{11.9}$), one Fe- ($Ca_{0.96}Cu_{2.85}Ti_{3.85}Fe_{0.018}O_{11.4}$) and one Ni-doped sample ($Ca_{0.97}Cu_{2.85}Ni_{0.022}Ti_{3.98}O_{11.8}$), the interatomic distances and the bond valence sums (BVS) were calculated from the structural results and are listed in Table I. The oxidation states are obtained by means of Brown's bond valence theory from the metal oxygen distances.[37] The structure has the specific particularity to form Cu square planes on the A-site of the primitive perovskite $ABO_3$. In fact, this square planar coordination of Cu by oxygen is formed by a large tilt of the $TiO_6$ octahedra with the consequence that the empty space for Ca cations is heavily reduced. This structural peculiarity seems to constitute a rather rigid lattice, barely susceptible for polar distortions as found in a variety of primitive perovskites. The Ca-O distance of about 2.6 Å, is significantly shorter than the expected ionic distance, 2.72 Å[38], and the Ca atoms are under compression with a formal valence of 2.11 instead of 2. Hence, the Ca ions are overbonded and the stretched Cu-O bond lengths lead to a decrease of the formal Cu valence to 1.79. The valence of Ti is close to 4. The bond valences as documented in Table I are in good agreement with data reported in literature.[39]

## IV. MAGNETIC PROPERTIES

The magnetic molar susceptibilities $\chi_m$ were measured in an applied magnetic field of 1000 Oe from 2 K to 400 K. Undoped CCTO shows antiferromagnetic (AFM) order with a Neel-temperature $T_N$ of about 25 K.[28,40] The Cu spins ($Cu^{2+}$, $d^9$, S = ½) order in ferromagnetic planes which are antiferromagnetically stacked along [111].[41] The molar magnetic susceptibilities $\chi_m$ as well as the inverse susceptibilities of ceramic CCTO doped with 1% Mn, 0.5% Fe, and 0.5% Ni are documented in Fig. 3 showing the typical signature of AFM ordering. Impurities in CCTO cause an increase in $\chi_m$, respectively a decrease in inverse $\chi_m$ (inset), below about 5 K. The paramagnetic behavior for temperatures above $T_N$ is fitted in a range from 100 K to 400 K with a Curie-Weiss law (the low temperature fractions of the Curie-Weiss laws are shown in the inset of Fig. 3). From these fits the Curie-Weiss temperatures $\Theta_{CW}$ and the paramagnetic moments have been determined. Values of $T_N$, $\Theta_{CW}$, and the effective magnetic moment $\mu_{eff}$ are listed in Table II. To compare these results with undoped polycrystalline CCTO, susceptibilities and inverse susceptibilities of differently tempered samples are shown in Fig. 4. Again the paramagnetic behavior above $T_N$ is fitted with a Curie-Weiss law in a temperature range from 100 K to 400 K. The obtained parameters are also listed in Table II.

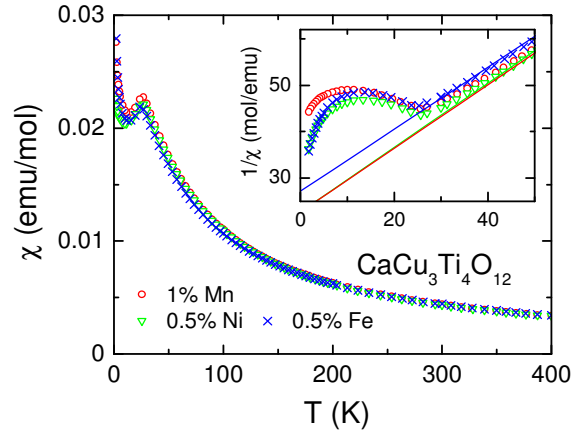

FIG. 3. (Color online) Temperature-dependent molar magnetic susceptibilities of polycrystalline CCTO doped with 1% Mn, 0.5% Ni, and 0.5% Fe. The inset shows the inverse magnetic susceptibilities at low temperatures. The lines are fits with a Curie-Weiss law performed at T > 100 K.

Undoped CCTO exhibits a simple collinear antiferromagnetic (AFM) structure below the magnetic ordering temperature. It results from the fact, that only the $Cu^{2+}$ ions with $3d^9$ configuration carry a local moment. $Ti^{4+}$ has $3d^0$ configuration and thus exhibits no magnetic moment. The magnetic structure of the perovskite-like compound CCTO can be explained by superexchange and spin-orbit coupling. Any direct exchange between $Cu^{2+}$ ions is negligibly small due to the large distances between these ions. Indirect exchange between the $S = ½$ magnetic moments of $Cu^{2+}$ are realized via $O^{2-}$ ions, via $Ti^{4+}$ ions with a triply degenerated $T_{2g}$ ground state,[28,31] or via $Ti^{4+}$ and $O^{2-}$ ions with five non-degenerated ground states.[28] It is generally believed that the superexchange path via the neighboring Ti ions is most important.[41,42] Grubbs et al.[28] reported that higher level (>3%) doping of $Fe^{3+}$ ($3d^5$ configuration) on the $Ti^{4+}$ site with a $3d^0$ configuration strongly affects the AFM transition temperatures. Table II and Fig. 3 of this work document that 0.5% Fe doping does not significantly affect the AFM ordering temperature $T_N$. The same holds for the CCTO samples slightly doped with manganese and nickel. In this work we assume that $Mn^{2+}$ and $Fe^{3+}$ ions are substituted on the $Ti^{4+}$ site. In literature it has been argued that $Ni^{2+}$ ions are more likely substituted for $Cu^{2+}$.[26] The AFM ordering temperatures of all compounds including polycrystalline samples prepared with different processing conditions are similar within experimental uncertainty. Based on our results and as documented in Table II an ordering temperature of 25.5(.1) K is the best estimate and we can state that at least marginal doping with Mn, Fe, or Ni does not influence the magnetic ordering temperature.

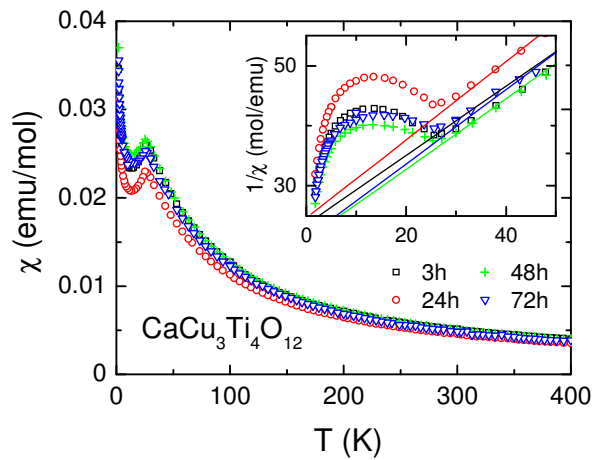

FIG. 4. (Color online) Temperature-dependent molar magnetic susceptibilities of polycrystalline CCTO tempered for 3 h, 24 h, 48 h, and 72 h. The inset shows the inverse magnetic susceptibilities at low temperatures. The lines are fits with a Curie-Weiss law performed at T > 100 K.

As outlined above, the values of the paramagnetic moments $\mu_{eff}$ and of the Curie-Weiss temperatures $\Theta_{CW}$ for doped and undoped CCTO were calculated from fits of $\chi(T)$ between 100 K and 400 K using a Curie-Weiss law. Similar to the findings of $T_N$ there seems to be no significant variation of the paramagnetic moments as documented in Table II, neither for the 3d doped samples nor for the polycrystalline materials synthesized using different tempering times. We find an average value of 3.55(15) $\mu_B$, which compares well with the expected paramagnetic moment per unit cell, taking the enhancement of the g-value for $Cu^{2+}$ into account. From the paramagnetic moment we deduce an effective g-value of 2.36, which seems to be strongly enhanced by spin-orbit coupling but which is not unlikely for Cu $3d^9$ systems.[42]

In contrast to the observed behavior of the ordering temperatures and the paramagnetic moments, there seems to be considerable sample synthesis and doping dependence of the Curie-Weiss temperatures. For undoped CCTO, the absolute values of $\Theta_{CW}$ decrease from 41 K to 34 K with increasing tempering time far beyond any experimental uncertainty. As known from literature (e.g., Ref. 7) the value of the CDC increases with tempering time. As reported in Ref. 32, the undoped CCTO samples investigated in our work also show an increase with tempering time and thus there seems to be a clear anti-correlation between Curie-Weiss temperature and enhancement of the dielectric constant. One can only speculate about the origin of this phenomenon. Tempering certainly increases the grain size.[7,32] It seems implausible that this could have an effect on the average exchange, which is measured by the Curie-Weiss temperature. Oxygen stoichiometry and homogeneity seem to be the most important parameters but within the experimental resolution neither oxygen stoichiometry nor homogeneity seem to be enhanced by increasing tempering time. As outlined above, superexchange in CCTO is mediated mainly via titanium ions and only to a much lesser extent via oxygen. Also from this point of view the strong influence of the tempering conditions on the mean exchange can not readily be understood. It is also unclear why significant changes in the average exchange as documented in Table II do not lead to concomitant changes in the magnetic ordering temperatures. In the doped compounds the absolute value of $\Theta_{CW}$ is rather low for Mn- and Ni-doping, but strongly enhanced for the Fe-doped samples. Indeed, there seems to be a very specific influence of Fe-doping on the magnetism of CCTO.

| Sample | $\Theta_{CW}$ / K | $T_N$ / K | $\mu_{eff}$ / f.u. |
| --- | --- | --- | --- |
| 1% Mn | 33.0 | 25.6 | 3.41 |
| 0.5% Ni | 32.7 | 25.6 | 3.39 |
| 0.5% Fe | 40.7 | 25.3 | 3.46 |
| 3h | 40.9 | 25.6 | 3.73 |
| 24h | 37.9 | 25.3 | 3.50 |
| 48h | 36.0 | 25.4 | 3.70 |
| 72h | 34.6 | 25.6 | 3.60 |

TABLE II. Absolute values for Curie-Weiss temperatures $\Theta_{CW}$, Neel-temperatures $T_N$, and effective magnetic moments per formula unit $\mu_{eff}$ for Mn, Fe, and Ni doped CCTO ceramics, as well as for undoped ceramics tempered for 3 h, 24 h, 48 h, and 72 h.

## V. DIELECTRIC PROPERTIES

Figure 5(a) and 5(b) show the temperature dependences of the dielectric constant and conductivity of polycrystalline CCTO doped with 1% Mn, measured for various frequencies in a broad temperature range from 4 K up to 425 K. At low frequencies and high temperatures a steplike increase of the dielectric constant from a value close to 100 to a CDC of the order of $10^4$ is observed, which can be ascribed to a Maxwell-Wagner (MW) relaxation originating from interface polarization. Similar behavior was also found in other CDC-materials.[43] The origin can either be internal interfaces, e.g., between conducting grains and insulating grain boundaries,[6,10,44] or, surface barrier layers, e.g., caused by a Schottky-type or metal-insulator-semiconductor diode.[20,11,21,22] In comparison to the temperature dependences of the dielectric properties of undoped CCTO, which are shown, e.g., in Ref. 32, the steplike behavior in $\varepsilon'(T)$ is shifted to significantly higher temperatures. The conductivity $\sigma'$ of the sample is related to the dielectric loss via $\sigma' = 2\pi\nu\varepsilon''\varepsilon_0$. Loss peaks arising at the points of inflection of the $\varepsilon'$ steps are typical relaxation features and, indeed, in Fig. 5(b) a corresponding peak shows up in the temperature-dependence of $\sigma'$ (e.g., for 1 Hz at 300 K). Because of the frequency dependence of the MW relaxation, the peaks in $\sigma'$ also show a characteristic shift with frequency.

to undoped CCTO[11,32] at comparable temperatures this feature is observed at much smaller frequencies. Within the MW picture, at high temperatures and/or low frequencies the dielectric properties are dominated by interfacial polarization. In contrast, at low temperatures and/or high frequencies, the high resistance of the insulating barriers becomes shorted by its capacitance and the intrinsic properties are detected. Thus, a steplike increase in $\sigma'(\nu)$ is expected as indeed observed in Fig. 6(b) (e.g., at around 100 Hz for the 400 K curve). The plateaus of $\sigma'(\nu)$ following this step (e.g., above 1 kHz for 400 K) are ascribed to the intrinsic dc conductivity. Lowering the temperature results in a shift of this dc plateau to smaller frequencies and in a decreasing value for the dc conductivity. This frequency-independent region also shows up in the temperature-dependent plot of Fig. 5(b) as a merging of the curves for different frequencies at the left flanks of the MW-relaxation peaks. The line in Fig. 5(b) indicates this region and is a direct measure of the temperature-dependent intrinsic conductivity of this material.

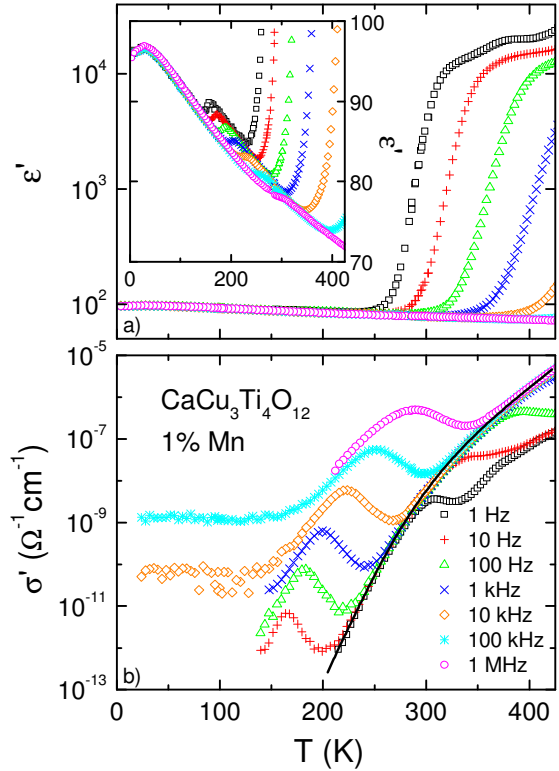

FIG. 5. (Color online) Temperature-dependent dielectric constant (a) and conductivity (b) of polycrystalline CCTO doped with 1% Mn at various frequencies. The inset shows $\varepsilon'(T)$ for temperatures below 300 K. The line in (b) provides an estimate of the intrinsic dc conductivity.

Figure 6(a) shows the frequency dependence of the dielectric constant for CCTO doped with 1% Mn. The MW relaxation also shows up in this representation: A steplike decrease of the dielectric constant from a value of $\varepsilon' \approx 2\times10^4$ at low frequencies and high temperatures to a high-frequency value of $\varepsilon' \approx 80$ is observed. In comparison

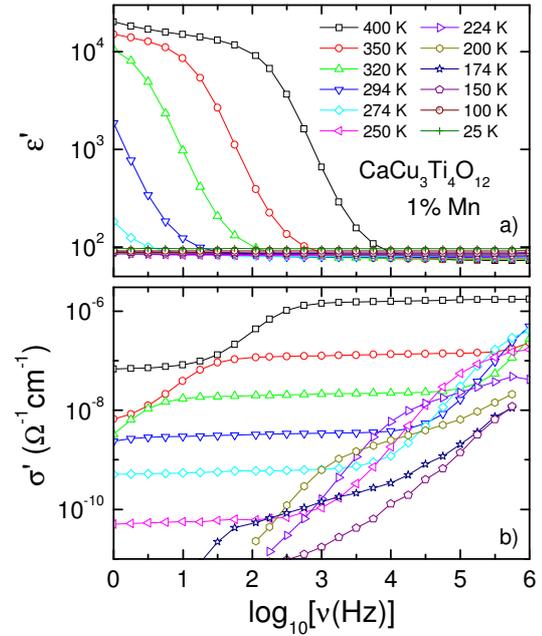

FIG. 6. (Color online) Frequency-dependent dielectric constant (a) and conductivity (b) of polycrystalline CCTO doped with 1% Mn at various temperatures. The lines are guides to the eyes.

We find that the dc conductivity of CCTO doped with only 1% Mn is reduced by more than 4 orders of magnitude at room temperature compared to undoped CCTO.[32] Obviously, manganese doping very effectively reduces the concentration of charge carriers, which in undoped CCTO ceramics are responsible for the dc conductivity.[27]

The reduced conductivity of the doped compound also explains the mentioned shifts of the relaxational features to higher temperatures (Fig. 5) or lower frequencies (Fig. 6) if compared to undoped CCTO: As discussed in detail, e.g., in Ref. 20, the relaxation time $\tau_{MW}$ of the MW relaxation in first approximation only depends on the bulk dc conductance $G_b$ (related to the conductivity via the geometry of the sample) and on the capacitance of the interface $C_c$, via $\tau_{MW} \approx C_c/G_b$. In this simplified picture it is obvious that a decrease of $G_b$ leads to an increase of $\tau_{MW}$. As $\tau_{MW} = 1/(2\pi\nu_p)$ with $\nu_p$ the frequency of the $\varepsilon'(\nu)$-step (or the $\varepsilon''(\nu)$-peak), the steplike increase in $\varepsilon'(\nu)$ (Fig. 6(a)) shifts to lower frequencies for lower $G_b$. Concerning the relaxation features in $\varepsilon'(T)$ and $\sigma'(T)$, one has to take into account the fact that the absolute value of $G_b(T)$ increases with increasing temperature (line in Fig. 5(b)). For the generally lower $G_b$ values in the doped samples, then the condition $\tau_{MW}(T) = 1/(2\pi\nu)$ at a given frequency $\nu$ is fulfilled only at higher temperatures if compared to CCTO.

One should note that concerning the applicability of these materials here a dilemma becomes obvious: Conductivities as low as possible certainly seem desirable for the construction of capacitive circuit elements. However, within the MW framework, lowering of the intrinsic conductivity (e.g., by doping as in the present case) reduces the temperature and frequency range where a CDC is observed. Thus, when optimizing materials with MW-generated CDCs, one should not aim at the reduction of the bulk conductivity. Instead, the conductivity of the insulating layers should be reduced, which has no influence on the relaxation time but dominates the overall conductance of the sample (at least at frequencies sufficiently low to observe the CDCs).

In addition to the MW-related peaks discussed above, in Fig. 5(b) further well-pronounced loss peaks are found at lower temperatures (e.g., at 200 K for 1 kHz). Thus, also a corresponding relaxation step should be present in $\varepsilon'(T)$. The inset of Fig. 5(a), showing a magnified view of the low-temperature region of $\varepsilon'(T)$, indeed reveals such a feature. The relaxation also shows up in Fig. 6(b) where shoulders in $\sigma'(\nu)$ (corresponding to peaks in $\varepsilon'' \sim \sigma'/\nu$) are observed at high frequencies. This relaxation most likely is of intrinsic nature. Up to now it could not be detected in undoped CCTO. Its origin is under debate: Either this small relaxation also exists in undoped CCTO but is superimposed by the strong Maxwell-Wagner relaxation or it is a specific feature of doped compounds.[28,30] We come back to this question when discussing similar relaxations in the Fe-doped compound.

In Fig. 5(b), at low temperatures $T < 150$ K the conductivity is nearly temperature independent. In this region, $\sigma'(\nu)$ shows a power-law increase (cf. curve for 150 K in Fig. 6(b)). Such a behavior is the typical signature of charge transport via hopping conductivity.[45]

In this class of materials not only the appearance of CDCs is of high scientific interest, but also the magnitude and temperature dependence of the intrinsic dielectric constant. Contrary to the undoped compound, in CCTO doped with 1% Mn the intrinsic dielectric constant can be detected at audio frequencies up to high temperatures, unobscured by the strong MW relaxation (inset of Fig. 5(a)). As mentioned above, compared to undoped CCTO in Mn-doped ceramics the MW relaxation is shifted towards higher temperatures and the intrinsic dielectric properties can be investigated in a broader temperature range. Leaving aside the small relaxation features discussed above, the inset of Fig. 5(a) reveals a significant increase of the intrinsic $\varepsilon'$ with decreasing temperature. In Ref. 31, based on measurements up to 1 GHz and in the infrared region, a similar increase was found for an undoped CCTO single crystal. It was ascribed to the softening of low lying transversal optical phonon modes and, in addition, to a concomitant increase of the effective ionic plasma frequencies of these modes. The increasing dielectric constant found for Mn-doped CCTO in Ref. 30 was ascribed to incipient ferroelectricity, which is a characteristic property of undoped $SrTiO_3$.[46,47,48] In $SrTiO_3$, relaxor-like ferroelectricity can be induced by small amounts of doping.[49] However, $SrTiO_3$ also can easily be driven into a ferroelectric polarized state by small external electric fields.[50] It seems that CCTO is not at all comparably susceptible, neither to doping nor to external electric fields.

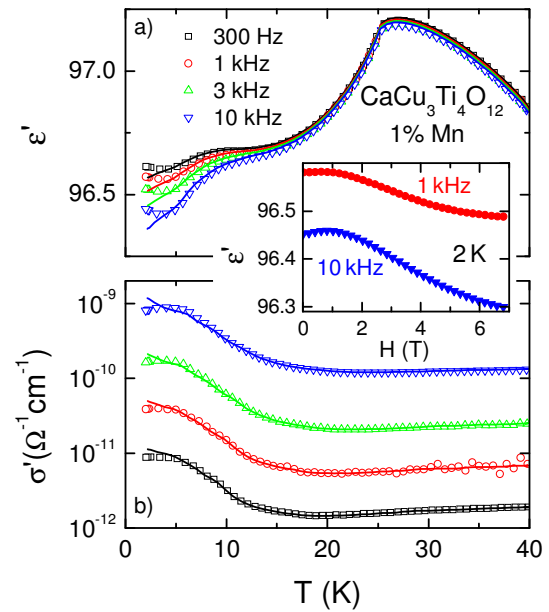

FIG. 7. (Color online) Temperature-dependent dielectric constant (a) and conductivity (b) of polycrystalline CCTO doped with 1% Mn. The measurements were performed at various frequencies in zero magnetic field (symbols) and with applied external magnetic field of 5 T (lines). The inset shows the magnetic-field dependence of $\varepsilon'$ at 2 K for two frequencies.

In the inset of Fig. 5 another interesting phenomenon is observed: At the AFM transition $\varepsilon'$ reveals a small but significant cusp in the temperature dependence, which was also reported in Refs. 25 and 28. Figure 7 shows the results of a measurement at low temperatures, using the high-resolution device AH2700a, which nicely reveals this feature. To investigate the magnetoelectric coupling in this compound in more detail, dielectric spectroscopy was additionally performed in an external magnetic field of $\mu_0 H = 5$ T (Fig. 7). The peak at the AFM transition is found to be frequency and magnetic-field independent (Fig. 7(a)). Its origin can be explained by a magnetic coupling of $Cu^{2+}$ and $Ti^{4+}$ ions via superexchange, which results in a stiffening of the lattice and a concomitant decrease of $\varepsilon'$ at the AFM transition. Focusing on the measurements with magnetic field, the dielectric constants show only slight deviations from those in zero field. Minor field-dependent effects appear only well below the AFM transition at $T < 10$ K, where the signature of a further, very weak relaxation step in $\varepsilon'(T)$ is found (Fig.

7(a)). In σ'(T) the corresponding relaxation peaks show up (Fig. 7(b)) also showing a slight dependence on the magnetic field. Such a low-temperature relaxational process at about 10 K was also reported for Nb- and Fe-doped samples.[28] Thus, it can be assumed to be a characteristic process, which may be also present in undoped CCTO. Its field-dependent behavior is also documented in the inset of Fig. 7 where ε'(H) is shown at 2 K. Obviously, the external magnetic field only slightly influences the second intrinsic relaxation.

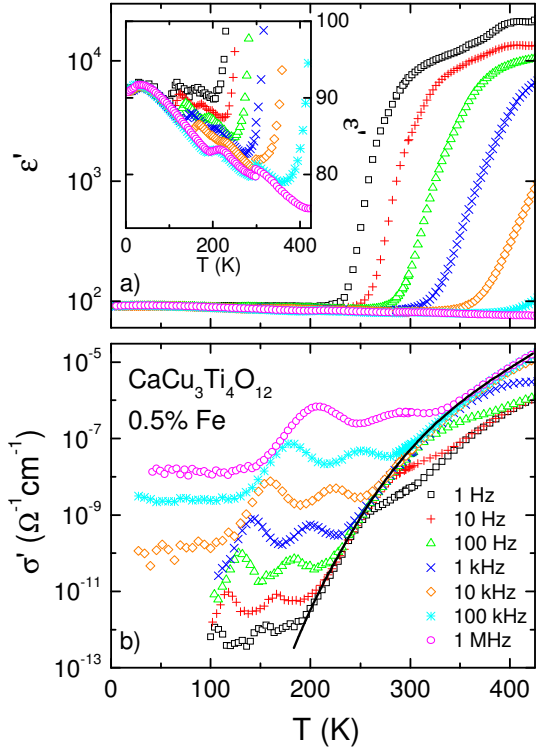

FIG. 8. (Color online) Temperature-dependent dielectric constant (a) and conductivity (b) of polycrystalline CCTO doped with 0.5% Fe, as measured at various frequencies. The inset shows ε' for temperatures below 300 K. The line in (b) provides an estimate of the intrinsic dc conductivity.

Figures 8(a) and 8(b) show the temperature dependences of the dielectric constant and conductivity of polycrystalline CCTO doped with 0.5% Fe, measured for various frequencies and temperatures from 4 K to 425 K. Like in the manganese-doped samples, a step like increase in the dielectric constant is observed at low frequencies and high temperatures. In comparison to undoped CCTO, the increase of the dielectric constant to colossal values higher than $\varepsilon' \approx 10^4$ again is shifted to higher temperatures. Modeling this behavior by a MW relaxation, as for the Mn-doped sample this observation can be explained by a decrease of the bulk dc conductivity (indicated by the line in Fig. 8(b)) by doping. Overall, the shifts of the MW relaxations compared to the undoped compound, observed in Figs. 5 and 8, clearly document the expected correlation between conductivity and Maxwell-Wagner relaxation.

Similar to the observation in Mn-doped compounds, the intrinsic dielectric constant increases with decreasing temperature (inset of Fig. 8(a)). Furthermore a small peak in the temperature dependence of the dielectric constant at the antiferromagnetic transition is present and compares well with the observation in Mn-doped compounds.

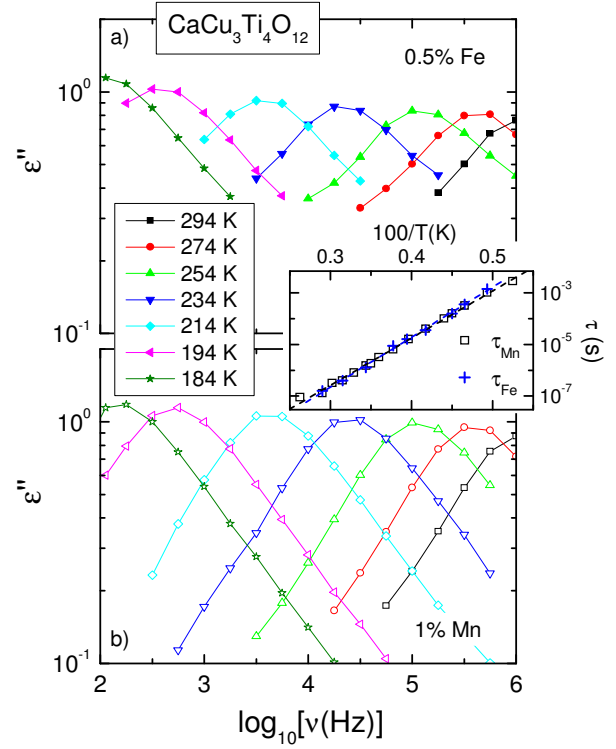

FIG. 9. (Color online) Frequency-dependent dielectric loss of polycrystalline CCTO doped with 0.5% iron (a) and with 1% manganese (b) at various temperatures. The figure focuses on the region of the intrinsic high-temperature relaxation. The lines connect the data points. The inset shows the relaxation times determined from the peak frequencies in Arrhenius representation with linear fits (dashed lines).

Compared to Mn-doped CCTO, there is one additional peak in the conductivity of the Fe-doped material (Fig. 8(b)) and, thus, two intrinsic relaxations are present. This result is consistent with that from Grubbs et al.[28] These relaxations also can be well identified by steps in the temperature-dependent dielectric constant as documented in the inset of Fig. 8(a). Comparing Figs. 5(b) and 8(b), it becomes obvious, that the intrinsic relaxation showing up at higher temperatures in Fe-doped CCTO seems to correspond to the single intrinsic relaxation observed in the Mn-doped sample (compare, e.g., the peaks of the 1 kHz curves both showing up at about 200 K in Figs. 5(b) and 8(b)). As documented in the literature, this intrinsic relaxation of Fe-doped CCTO is also present in compounds with significantly higher Fe content.[28] This observation points towards the fact that the second lower-temperature intrinsic relaxation originates from Fe-induced dipolar defects, while the higher-temperature relaxation may be present in all samples, either undoped or doped. In the undoped compounds this relaxation probably is hidden underneath the giant Maxwell-Wagner relaxation.

To investigate these intrinsic relaxations in more detail, Fig. 9 documents the frequency dependence of the dielectric loss in the relevant temperature region for both doped compounds. Typical relaxation peaks show up shifting through the investigated frequency window with changing temperature. The relaxations in both compounds behave rather similar: The loss peaks are located at comparable frequencies and both relaxations are of similar amplitude. The inset of Fig. 9 shows the temperature dependence of the relaxation times $\tau$, determined via $\tau \propto 1/\nu_p$, in an Arrhenius representation. It provides clear experimental evidence that both relaxations follow a nearly identical activated behavior, $\tau = \tau_0 \exp[E_b/(k_B T)]$. In this representation, the energy barrier $E_b$ and prefactor $\tau_0$ are revealed by linear fits of $\tau(T)$. From $\tau_0$ an attempt frequency can be calculated via $\nu_0 = 1/(2\pi\tau_0)$. The values obtained for Fe-doped CCTO are $E_b(\text{Fe}) = 0.39$ eV and $\nu_0(\text{Fe}) = 4.6$ THz and those for Mn-doped CCTO are $E_b(\text{Mn}) = 0.36$ eV and $\nu_0(\text{Mn}) = 1.5$ THz. The attempt frequencies of Fe- and Mn-doped CCTO are in the range of low lying phonon modes.[12,31] The similarities of the Arrhenius parameters found for the relaxations of both compounds again indicate that they can be ascribed to the same origin.

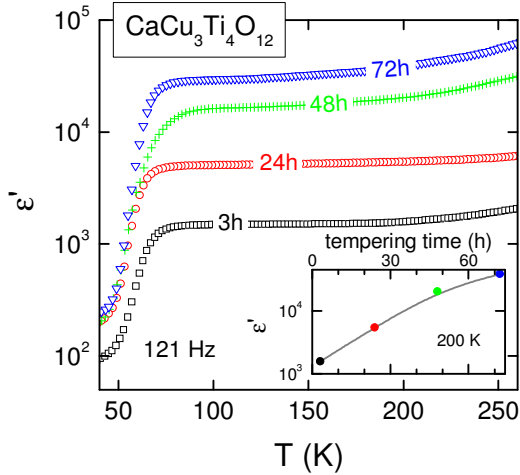

FIG. 10: (Color online) Temperature-dependent dielectric constants of undoped polycrystalline CCTO at 121 Hz for samples subjected to different tempering times. The inset shows the values of the CDC as function of tempering time. The line is drawn to guide the eyes.

Now we will briefly discuss the tremendous influence of the conditions of synthesis on the dielectric and conducting properties of CCTO. Fig. 10 provides experimental evidence for the strong impact of tempering time on the value of the CDCs. An increase of tempering time from 3 h to 72 h enhances the dielectric constants almost by a factor of 200. However, interestingly the temperature location of the MW-derived step does not vary significantly.

The inset of Fig. 10 documents the increase of the dielectric constant vs. tempering time. Close to 100 hours, saturation in the magnitude of the dielectric constant can be expected. It is most likely that this increase in $\varepsilon'$ exclusively results from an increase of grain size within the ceramics. Two different effects may explain this finding: The reduction in the number of internal barrier layers, which are due to grain boundaries,[7,51] or the increase of the smoothness of the surface of the ceramics caused by the increasing grain size.[32] An improved smoothness of the sample surface allows better "wetting" of the metal contact, i.e. the area of direct metal-semiconductor contact increases and consequently a more perfect development of Schottky-type diodes arises.[11,21]

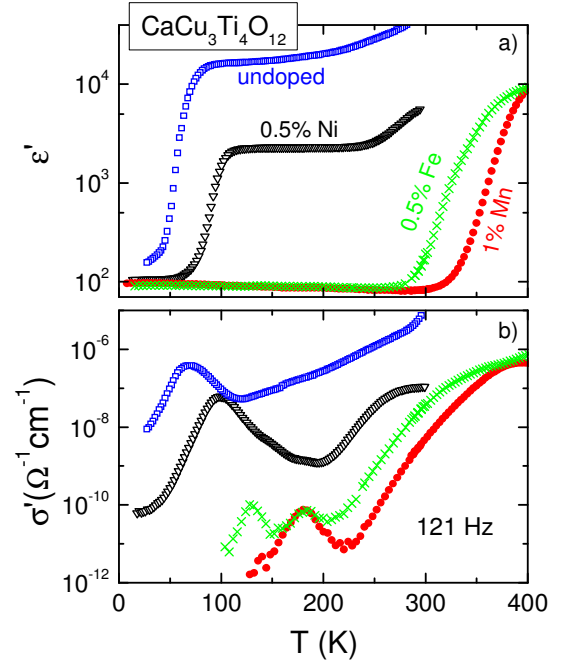

FIG. 11: (Color online) Temperature-dependent dielectric constants (a) and conductivities (b) of undoped and doped (1% Mn, 0.5% Fe, and 0.5% Ni) polycrystalline CCTO at 121 Hz.

Now we come back to the dielectric properties of the doped CCTO samples. Figure 11 compares the temperature dependences of the dielectric constants (a) and conductivities (b) at 121 Hz of undoped and doped CCTO. Undoped CCTO shows the well-known step like increase, reaching $\varepsilon' \approx 17000$. The onset of a second, non-intrinsic high-temperature relaxation, which has been described previously,[21] leads to the additional increase of $\varepsilon'(T)$ above about 200 K. These two relaxations can be ascribed to MW effects originating from two interface barriers, e.g., a metal-insulator-semiconductor diode or the combination of surface and internal barrier capacitances.[32] As a consequence of the substitution of 3d ions even in marginal (less than 1%) concentrations, the behavior of the dielectric constant and conductivity changes drastically. $\text{Ni}^{2+}$ doping (most likely isoelectronic substitution on the copper-place) shifts the Maxwell-Wagner step to somewhat higher temperatures. The value of the CDC decreases, but this phenomenon may mainly result from the grain size and thus the CDCs can hardly be compared in the doped ceramics. The onset of a second relaxation is shifted to higher temperature, but still is present. In Mn- and Fe-doped CCTO the step-like increase to colossal values of the dielectric constant appears at much higher temperatures, close to room temperature.

This raises some doubts on the interpretation that nanoscale disorder of CCTO, namely Ca and Cu atomic substitution, is responsible for the CDCs.[52] It seems not reasonable to assume that 0.5% of manganese or 1% of iron can effectively influence this nanoscale disorder. As demonstrated in Fig. 11, $\varepsilon'(T)$ of the Mn- and Fe-doped samples is dominated by the bulk (intrinsic) dielectric properties almost up to room temperature (when measured at 121 Hz) and even up to 425 K at 1 MHz. The intrinsic dc conductivity, which can be read off at the low-frequency flank of the MW-relaxation peaks in Fig. 11(b) (cf. lines in Figs. 5(b) and 8(b)), decreases by orders of magnitude in the series undoped, Ni-, Fe-, and Mn-doped. This is directly related to the shift of the relaxation steps revealed by Fig. 11(a). Assuming that the conductivity in CCTO mainly depends on slight oxygen deficiencies, it seems that the free charge carriers are nearly completely compensated by non-stoichiometric doping with manganese or iron.

## VI. SUMMARY AND CONCLUSIONS

In the present work we have provided a detailed structural, magnetic, and dielectric characterization of CCTO ceramics doped with small amounts of 3d ions, like Fe, Mn, or Ni and of undoped CCTO samples subjected to different processing conditions. This work has been undertaken in order to find experimental evidence for the influence of synthesis and substitution on the dielectric properties. But we also wanted to establish possible correlations between structural details (lattice constants, bond lengths, etc.), magnetic, and dielectric properties. From this work it follows that structural details only result in minor effects on the dielectric as well as magnetic properties. For example, the magnetic ordering temperature and the effective moment are rather constant for all ceramics investigated (see Table II). However, there seems to be a dramatic influence of tempering time as well as of doping on the Curie-Weiss constant, which can hardly be understood.

By dielectric spectroscopy we have provided detailed information on the temperature and frequency dependence of the dielectric constant, loss, and conductivity. One main result of this work is that the intrinsic conductivity is dramatically influenced by doping and so is the Maxwell-Wagner relaxation, which crucially depends on the bulk conductivity of the material under investigation. Another important finding is that in Fe- and Mn-doped CCTO ceramics the intrinsic dielectric constant can be observed almost up to room temperature, even at low measuring frequencies. This is nicely documented in Figs. 5(a) and 8(a). On the other hand, tempering, which mainly influences the grain size of the ceramics, leads to a strong variation of the value of the CDC without having any considerable impact on the temperature or frequency location of the MW relaxation.

We provide clear indications that intrinsic relaxations can be observed in the doped ceramics. We document that one relaxation is nearly identical for the iron- and manganese-doped compounds. We speculate that this relaxation can be observed in all ceramics, but usually is hidden under the huge MW-relaxation. In the Fe-doped ceramic, we have detected a second intrinsic relaxation, which obviously results from the Fe ions. Furthermore, high-precision measurements in the Mn-doped sample reveal the presence of a relaxation at very low temperatures, T < 10 K, which also seems to be a universal feature of CCTO. Finally, we detected a significant influence of the AFM ordering on the dielectric constant (Fig. 7a) and found a minor dependence of the dielectric constant on external magnetic field at low temperatures, T < 10 K.

## ACKNOWLEDGMENTS

This work was supported by the Commission of the European Communities, STREP: NUOTO, NMP3-CT-2006-032644 and by the DFG via the SFB 484. JL would like to thank the China Scholarship Council for the support of his oversea Ph.D. studies. We thank Dana Vieweg for performing SQUID and X-Ray analyses.

[1] B. Bochu, M. N. Deschizeaux, J.C. Joubert, A. Collomb, J. Chenavas, and M. Marezio, J. Solid State Chem. **29**, 291 (1979).
[2] C. C. Homes, T. Vogt, S. M. Shapiro, S. Wakimoto, and A. P. Ramirez, Science **293**, 673 (2001).
[3] M. A. Subramanian, D. Li, N. Duan, B. A. Reisner, and A. W. Sleight, J. Solid State. Chem. **151**, 323 (2000).
[4] N. A. Vasil'ev, and O. S. Volkova, Low Temp. Phys. **33**, 895 (2007).
[5] A. P. Ramirez, M. A. Subramanian, M. Gardel, G. Blumberg, D. Li, T. Vogt, and S. M. Shapiro, Solid State Commun. **115**, 217 (2000).
[6] D. C. Sinclair, T. B. Adams, F. D. Morrison, and A. R. West, Appl. Phys. Lett. **80**, 2153 (2002).
[7] T. B. Adams, D. C. Sinclair, and A. R. West, Adv. Mat. **14**, 1321 (2002).
[8] M. A. Subramanian and A. W. Sleight, Solid State Sci. **4**, 347 (2002).
[9] M. H. Cohen, J. B. Neaton, L. He, and D. Vanderbilt, J. Appl. Phys. **94**, 3299 (2003).
[10] S.-Y. Chung, I.-D. Kim, and S.-J. L. Kang, Nature Mat. **3**, 774 (2004).
[11] P. Lunkenheimer, R. Fichtl, S. G. Ebbinghaus, and A. Loidl, Phys. Rev. B **70**, 172102 (2004).
[12] L. He, J. B. Neaton, M. H. Cohen, D. Vanderbilt, and C. C. Homes, Phys. Rev. B **65**, 214112 (2002).
[13] J. Sebald, S. Krohns, P. Lunkenheimer, S. G. Ebbinghaus, S. Riegg, A. Reller, and A. Loidl, arXiv:0903.4189
[14] R. Weht, and W. E. Pickett, Phys. Rev. B **65**, 014415 (2002).
[15] M. N. Deschizeaux, J. C. Joubert, A. Vegas, A. Collomb, J. Chenavas, and M. Marezio, J. Solid State Chem. **19**, 45 (1976).
[16] J. Chenavas, J. C. Joubert, M. Marezio, and B. Bochu, J. Solid State Chem. **14**, 25 (1975).
[17] W. Kobayashi, I. Terasaki, J. Takeya, I. Tsukada, and Y. Ando, J. Phys. Soc. Jpn. **73**, 2373 (2004).
[18] A. Krimmel, A. Günther, W. Kraetschmer, H. Dekinger, N. Büttgen, A. Loidl, S. G. Ebbinghaus, E.-W. Scheidt, and W. Scherer, Phys. Rev. B **78**, 165126 (2008).
[19] see, e.g., Y. Lin, Y. B. Chen, T. Garret, S. W. Liu, C. L. Chen, L. Chen, R. P. Bontchev, A. Jacobson, J. C. Jiang and E. I. Meletis, Appl. Phys. Lett. **81**, 631 (2002); R. Lo Nigro, R. G. Toro, G. Malandrino, I. L. Fragala, M. Losurdo, M. M. Giangregorio, G. Bruno, V. Raineri, and P. Fiorenza, J. Phys. Chem. B **110**, 17460 (2006); P. Fiorenza, R. Lo Nigro, A. Sciuto, P. Delugas, V. Raineri, R. G. Toro, M. R. Catalano, and G. Malandrino, J. Appl. Phys. **105**, 061634 (2009).
[20] P. Lunkenheimer, V. Bobnar, A. V. Pronin, A. I. Ritus, A. A. Volkov, and A. Loidl, Phys. Rev. B **66**, 052105 (2002).
[21] S. Krohns, P. Lunkenheimer, S. G. Ebbinghaus, and A. Loidl, Appl. Phys. Lett. **91**, 022910 (2007); *ibid*. **91**, 149902 (2007).

[22] G. Deng, T. Yamada, and P. Muralt, Appl. Phys. Lett. **91**, 202903 (2007).

[23] V. Brize, G. Gruener, J. Wolfman, K. Fatyeyeva, M. Tabellout, M. Gervais, and F. Gervais, Mat. Sci. Eng. B **129**, 135 (2006).

[24] M. Maglione, J. Phys. Condens. Matter **20**, 322202 (2008).

[25] W. Kobayashi, and I. Terasaki, Physica B **329**, 771 (2003).

[26] S.-Y. Chung, S.-Y. Choi, T. Yamamoto, Y. Ikuhara, and S.-J. L. Kang, Appl. Phys. Lett. **88**, 091917 (2006).

[27] M. Li, A. Feteira, D. C. Sinclair, and A. R. West, Appl. Phys. Lett. **88**, 232903 (2006).

[28] R. K. Grubbs, E. L. Venturini, P. G. Clem, J. J. Richardson, B. A. Tuttle, and G. A. Samara, Phys. Rev. B **72**, 104111 (2005).

[29] V. Brize, C. Autret-Lambert, J. Wolfman, M. Gervais, P. Simon, and F. Gervais, Solid State Science **11**, 875 (2009).

[30] M. Li, A. Feteira, D. C. Sinclair, and A. R. West, Appl. Phys. Lett. **91**, 132911 (2007).

[31] Ch. Kant, T. Rudolf, F. Mayr, S. Krohns, P. Lunkenheimer, S. G. Ebbinghaus, and A. Loidl, Phys. Rev. B **77**, 045131 (2008).

[32] S. Krohns, P. Lunkenheimer, S. G. Ebbinghaus, and A. Loidl, J. Appl. Phys. **103**, 084107 (2008).

[33] A. Douy, Intern. J. Inorg. Mat. **3**, 699 (2001).

[34] A. Hassini, M. Gervais, J. Coulon, V. T. Phuoc, and F. Gervais, Mat. Sci. Eng. B **87**, 164 (2001).

[35] U. Schneider, P. Lunkenheimer, A. Pimenov, R. Brand, and A. Loidl, Ferroelectrics **249**, 89 (2001).

[36] J. Rodriguez-Carvajal, Physica B **192**, 55 (1993).

[37] I. D. Brown in *Structure and bonding in crystals*, M. O'Keefe, A. Navrotsky, Eds Academic Press: New York 2, 1-30 (1981).

[38] R. D. Shannon, Acta Crystallogr. A **32**, 751 (1976).

[39] J. Li, M. A. Subramanian, H. D. Rosenfeld, C. Y. Jones, B. H. Toby, A. W. Sleight, Chem. Mater. **16**, 5223 (2004).

[40] A. Collomb, D. Samaras, B. Bochu, and J. C. Joubert, Phys. Stat. Sol. A **41**, 459 (1977).

[41] Y. J. Kim, S. Wakimoto, S. M. Shapiro, P. M. Gehring, A. P. Ramirez, Solid State Commun. **121**, 625 (2002).

[42] C. Lacroix, J. Phys. C: Solid St. Phys. **13**, 5125 (1980).

[43] T. Park, Z. Nussinov, K. R. A. Hazzard, V. A. Sidorov, A. V. Balatsky, J. L. Sarrao, S.-W. Cheong, M. F. Hundley, J.-S. Lee, Q. X. Jia, and J. D. Thompson, Phys. Rev. Lett. **94**, 017002 (2005); P. Lunkenheimer, T. Götzfried, R. Fichtl, S. Weber, T. Rudolf, A. Loidl, A. Reller, and S.G. Ebbinghaus, J. Solid State Chem. **179**, 3965 (2006); S. Krohns, P. Lunkenheimer, Ch. Kant, A.V. Pronin, H.B. Brom, A.A. Nugroho, M. Diantoro, and A. Loidl, Appl. Phys. Lett. **94**, 122903 (2009); B. Renner, P. Lunkenheimer, M. Schetter, A. Loidl, A. Reller, and S. G. Ebbinghaus, J. Appl. Phys. **96**, 4400 (2004); J. L. Cohn, M. Peterca, and J. J. Neumeier, J. Appl. Phys. **97**, 034102 (2005).

[44] P. Fiorenza, R. Lo Nigro, C. Bongiorno, V. Raineri, M. C. Ferarrelli, D. C. Sinclair, and A. R. West, Appl. Phys. Lett. **92**, 182907 (2008).

[45] S. R. Elliott, Adv. Phys. **36**, 135 (1987); A. R. Long, Adv. Phys. **31**, 553 (1982).

[46] K. A. Müller, and H. Burkard, Phys. Rev. B **19**, 3593 (1979).

[47] J. H. Barrett, Phys. Rev. **86**, 118 (1952).

[48] J. Hemberger, P. Lunkenheimer, R. Viana, R. Böhmer, and A. Loidl, Phys. Rev. B **52**, 13159 (1995).

[49] J. G. Bednorz, and K. A. Müller, Phys. Rev. Lett. **52**, 2289 (1984); C. Ang, Z. Yu, P. Lunkenheimer, J. Hemberger, and A. Loidl, Phys. Rev. B **59**, 6670 (1999).

[50] J. Hemberger, M. Nicklas, R. Viana, P. Lunkenheimer, A. Loidl, and R. Böhmer, J. Phys.: Cond. Matter **8**, 4673 (1996).

[51] J. L. Zhang, P. Zheng, C. L. Wang, M. L. Zhao, J. C. Li, and J. F. Wang, Appl. Phys. Lett. **87**, 142901 (2005).

[52] Y. Zhu, J. C. Zheng, L. Wu, A. I. Frenkel, J. Hanson, P. Northrup, and W. Ku, Phys. Rev. Lett. **99**, 037602 (2007).